\documentclass[11pt,a4paper]{article}
\pdfoutput=1
\usepackage{jheppub}

\usepackage{float}
\usepackage{amsmath}
\usepackage{cleveref}
\usepackage{xcolor}

\newcommand{\ud}{\mathrm{d}}
\newcommand{\AI}{\mathcal{I}}

\crefname{equation}{eqn.}{eqns.}
\Crefname{equation}{Eqn.}{Eqns.}
\crefname{figure}{Fig.}{Figs.}
\Crefname{figure}{Fig.}{Figs.}

\title{ Quantum tunnelling, real-time dynamics and Picard-Lefschetz thimbles.
}

\author[b]{Zong-Gang Mou,}
\author[a]{Paul M. Saffin,}
\author[b]{Anders Tranberg,}

\affiliation[a]{School of Physics and Astronomy, University Park, University of Nottingham,\\ Nottingham NG7 2RD, United Kingdom}
\affiliation[b]{Faculty of Science and Technology, University of Stavanger, 4036 Stavanger, Norway}

\emailAdd{zonggang.mou@uis.no}
\emailAdd{paul.saffin@nottingham.ac.uk}
\emailAdd{anders.tranberg@uis.no}

\abstract{
We follow up the work, where in light of the Picard-Lefschetz thimble approach, we split up the real-time path integral into two parts: the initial density matrix part which can be represented via an ensemble of initial conditions, and the dynamic part of the path integral which corresponds to the integration over field variables at all later times. This turns the path integral into a two-stage problem where, for each initial condition, there exits one and only one critical point and hence a single thimble in the complex space, whose existence and uniqueness are guaranteed by the characteristics of the initial value problem. In this paper, we test the method for a fully quantum mechanical phenomenon, quantum tunnelling in quantum mechanics. We compare the method to solving the Schr\"odinger equation numerically, and to the classical-statistical approximation, which emerges naturally in a well-defined limit. We find that the Picard-Lefschetz result matches the expectation from quantum mechanics and that, for this application, the classical-statistical approximation does not.
}

\begin{document}

\maketitle

\section{Introduction}
\label{sec:Intro}

The computation of real-time correlators in quantum mechanics may be formally phrased through a path integral formulation of the time evolution of the density matrix. Given that a correlator is expressed through
\begin{align}\label{eq:expectation_O_v1}
\langle\hat{\mathcal{O}}(t)\rangle=\textrm{Tr}[\rho_0 \hat{\mathcal{O}}(t)],
\end{align}
one may rewrite this as a path integral 
\begin{align}
\label{eq:pathintegral}
\langle\hat{\mathcal{O}}(t)\rangle =\frac{1}{Z}\int {\mathcal D}\phi \rho_0\, e^{\frac{i}{\hbar}\int_{\mathcal C} \ud t L}\mathcal{O}, \qquad Z=\int {\mathcal D}\phi \rho_0 e^{\frac{i}{\hbar}\int_{\mathcal C} \ud t L}
\end{align}
The operator $\hat{\mathcal{O}}$ is assumed to be expressed in terms of some fields $\phi$, that now live on the Schwinger-Keldysh contour $\mathcal{C}$ \cite{Schwinger:1960qe,Keldysh:1964ud} in the complex time-plane. This means that their time index starts at the initial time $t=0$, runs to some late time, and returns to $t=0$ (see below). $\rho_0$ is a representation of the initial density matrix, $L$ is the Lagrangian expressed in terms of the fields $\phi$ and $Z$ is the partition function.

This integral can in certain simple cases be computed explicitly, and advanced perturbative techniques allows the computation of real-time correlators, say in equilibrium. However, in the cases where non-perturbative or out-of-equilibrium information is of interest, the situation becomes very challenging. Numerical simulations must then often be employed, and since the integrand is a highly oscillatory complex phase (the Lagrangian $L$ is usually real), convergence of numerical integration is in general poor due to this ``sign problem".

There are a number of ways to try and sidestep this problem. One is to solve for the time evolution of the fields, or correlators of the fields, through some effective, systematically truncated, evolution equations. Good examples include the classical statistical approximation, where an ensemble mimicking the quantum initial density matrix are evolved using straightforward classical equations of motion. In cases where the occupation numbers are large, and the system is far from thermal equilibrium, this is a powerful and easily implementable approximation (see for instance \cite{GarciaBellido:2002aj,Smit:2002yg,Aarts:2001yn,Berges:2013lsa,Mou:2017xbo}). Truncated Schwinger-Dyson equations for two- or higher-point correlators have been used to good effects for the approach to equilibrium, as well as in cosmological applications involving scalar and fermion fields (see \cite{Berges:2000ur,Berges:2002wr,Arrizabalaga:2004iw,Arrizabalaga:2005tf}). Stochastic equations have also proven a valuable tool for concrete cases, that allow for a mapping of the full dynamics to such a description in a systematic way (see for instance \cite{Bodeker:1998hm,DOnofrio:2014rug}). 

In some cases, however, one must return to the direct computation of the path integral (\ref{eq:pathintegral}), but standard Monte-Carlo techniques, routinely and effectively applied to euclidean systems in QCD, fall short. A number of ways to try and resolve this sign problem have been proposed. Stochastic quantization is particularly promising, providing a Langevin type equation to sample the field space \cite{Aarts:2008rr,Aarts:2009uq,Sexty:2013ica}. For a real-time path integral, the procedure leads to complex Langevin equations, driving the field $\phi$ away from the real axis sampling the entire complex plane. This effective doubling of the degrees of freedom helps to significantly improve convergence and one may show that under sensible assumptions, the correct physical result is achieved .

Another remedy for the sign problem is to again complexify the real integration variables $\phi$, but rather than doubling the dimensionality of field space $R^n\rightarrow C^n$, we constrain the field variables to live on a specific manifold in the complex plane of real dimension $n$. The task is then to find a manifold where the integral is better behaved, and we can expect convergence with a manageable amount of importance sampling. This manifold could be the Lefschetz thimbles \cite{Cristoforetti:2012su,Cristoforetti:2013wha,Mukherjee:2013aga,Behtash:2015kna,Tanizaki:2015rda,Tanizaki:2017yow}, or other manifolds such as generalized thimbles \cite{Alexandru:2015xva,Alexandru:2015sua,Alexandru:2016gsd,Alexandru:2017oyw,Alexandru:2017lqr,Alexandru:2017czx,Alexandru:2018fqp,Alexandru:2018ngw,Fukuma:2017fjq,Ulybyshev:2019hfm,Ulybyshev:2019fte}, and in each case, we need to provide a numerical algorithm to find the given manifold. The upshot is that as a consequence of Cauchy's theorem (generalised to $n$ dimensions), the path integral on any sensible deformation of $R^n$ will give the same result.

Since the sampling is performed along the thimble, the method works best when the thimble is unique, or when there is a way of including each thimble systematically one by one. In thermal equilibrium \cite{Das:1997gg}, it is customary to include the initial density matrix $\rho_0$ in the path integral. This leads to the complex time-contour being extended to $t=-i/T$, with $T$ the temperature. Further, the field configurations are prescribed to be periodic in the time variable. In this case, the Monte Carlo algorithm will need to sample the entire configuration space, including all the thimbles,. 

In \cite{Mou:2019tck}, we argued that for initial value problems, one should instead think of the initial density matrix as an ensemble of initial conditions, very similar to the classical statistical approximation. Each realisation of the initial density matrix leads to a classical trajectory, and each such trajectory is a ``critical" configuration corresponding to one unique thimble. Hence, to the extent that the initial state permits it\footnote{This initial could be Gaussian, or in some other distribution that is easy to sample.}, all the thimbles will be included in the total path integral by sampling the ensemble of initial conditions one by one. In \cite{Mou:2019tck} we implemented this in quantum mechanics (0-dimensional field theory) for a simple scalar model, and computed the non-equal time correlator. 

Still, it would be interesting to stress-test our method for a truly quantum mechanical phenomenon, in a non-trivial system.
And so in the present paper we study quantum mechanical tunnelling (for applications of Lefschetz thimble to tunnelling problem from a more theoretical viewpoint, see \cite{Tanizaki:2014xba,Cherman:2014sba,Ai:2019fri}), precisely because we expect this process to challenge the numerical path integral.
Furthermore, we expect the difference between the thimble approach and the classical-statistical approximation to be much more pronounced. In our real-time thimble approach, we sample the initial conditions and solve for the classical trajectories. The further identification and sampling along the thimbles provides the quantum contribution, but omitting this step allows us to recover the standard classical-statistical averaging. This allows a straightforward comparison between the two methods.

The structure of the paper is as follows:
In \cref{sec:path_integral}, we present the closed-time path integral in terms of the initial density matrix and thimble.
In \cref{sec:tunnel}, we solve the textbook quantum mechanical tunnelling problem of ground state tunnel splitting, utilizing different methods.
Finally, \cref{sec:conclu} is our summary and conclusions.

\section{Setting up the path integral}
\label{sec:path_integral}

The section is devoted to a short, but hopefully sufficiently self-contained, description of the closed-time path integral and thimble methods.
For a thorough discussion of the topics, we refer the reader to \cite{Mou:2019tck}.
Since  quantum field theory problems are what we ultimately want to tackle, we keep a spatial index in this section.
We have in mind a single real scalar field with a double-well potential,
\begin{align}\label{eq:V}
{\mathcal V}(\phi) &= \frac{m^2}{2}\phi^2-g_3\phi^3+\frac{\lambda}{24}\phi^4,
\end{align}
which we will use later in the next section. All the conclusions drawn in the present section, however, hold true for arbitrary potential.

\subsection{The closed-time path integral}
\label{sec:separation}

\begin{figure}[t]
\centering
\includegraphics[width=0.65\textwidth]{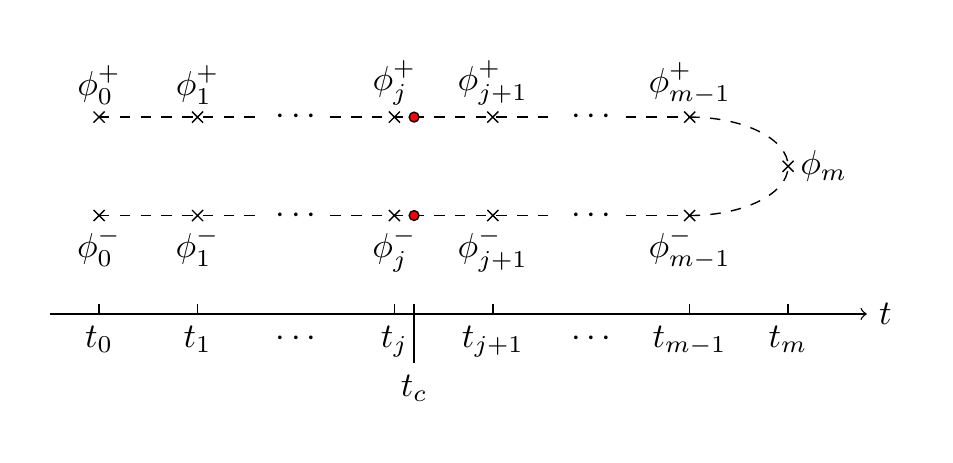}
\caption{Closed time path, ${\cal C}$, used for solving the real-time path integral.}
\label{fig:CTP}
\end{figure}

As described briefly above, the Schwinger-Keldysh closed-time path integral \cite{Schwinger:1960qe,Keldysh:1964ud} appears naturally when we evaluate correlation functions at time $t$, given an initial density matrix at $t_0$,
\begin{align}\label{eq:expectation_O}
\langle\hat{\cal O}(t)\rangle&={\rm Tr}\left( \hat{\cal O}(t)\hat\rho(t_0) \right)
={\cal N}\int{\cal D}\phi\; {\cal O}(t)\langle\phi^+_0;t_0|\hat\rho|\phi^-_0;t_0\rangle e^{ \frac{i}{\hbar}\int_{\cal C}\ud t L }.
\end{align}
As shown in \cref{fig:CTP}, the path comprises two branches starting at some initial time $t_0$ (typically, $t_0=0$), and ending at some final time $t_m$. The upper branch hosts fields $\phi^+_j$, while the lower one hosts $\phi^-_j$. In the following, we will think of a finite set of discrete values of time $t$ separated by $\ud t$, as appropriate for the numerical simulations to be performed below. Formally, the continuum limit arises from taking $\ud t\rightarrow 0$.
These branches appear when one inserts complete sets of states\footnote{
These states are eigenstates of the operator $\hat\phi(t_j)$.
One may reach the same result by using the unitary time-evolution operator.
}, $|\phi_j,t_j\rangle$, into (\ref{eq:expectation_O_v1}), located at times from $t_0$ to some time $t_m$, later than any time appearing in the operator $\hat{\cal O}$. 
One then finds a number of inner products of the form $\langle\phi_i,t_i|\phi_{j},t_{j}\rangle$, which may be evaluated by using the Hamiltonian to evolve each state to some intermediate time $t_c$. In \cite{Mou:2019tck} we took $t_c$ to lie at the midpoint between $t_i$ and $t_j$.
Besides the choice of midpoint, one can also choose the forward or backward difference, for instance, to select $t_j$ or $t_i$ to be $t_c$.
But this comes with a warning.
For instance, if one chooses the forward difference on the upper branch, i.e., evaluating $\langle \phi^+_{j+1},t_{j+1} |\phi^+_{j},t_{j}\rangle$ at $t_{j+1}$, then one should also evaluate the lower branch amplitude $\langle \phi^-_{j},t_{j} | \phi^-_{j+1},t_{j+1} \rangle$ at $t_{j+1}$, even though this counts as a backward difference.
In particular, the expressions on the upper and lower branches should be the complex-conjugate of each other.
This requirement is obvious when we calculate the expectation value of $\phi_m$ in Quantum Mechanics,
\begin{align}\label{eq:wavefunction}
\langle \hat \phi_m \rangle \equiv \int \ud \phi_m \Psi^* \phi_m \Psi,
\end{align}
given the wavefunction $\Psi=\int^{\phi_m} {\mathcal D}\phi\exp\left(\frac{i}{\hbar}\int_{t_0}^{t_m}\ud t L\right)\Psi_{0}$.
As a bonus of \cref{eq:wavefunction}, the complex-conjugation enforces all $\phi_m$ terms in the exponent to vanish, except the temporal derivative terms, since the Lagrangian is real.
This conclusion also holds true for quantum field theory. That is to say, as long as $t_c$ is chosen consistently, the exponent in (\ref{eq:expectation_O}) does not contain $\phi_m$ terms other than through  $\phi_m\phi_{m-1}^{\pm}$.
This linear dependence on $\phi_m$ is crucial in what follows.
\footnote{
We have inserted only a single $\phi_m$ field at $t_m$, which is manifested in \cref{eq:wavefunction}.
Actually, even if we insert both upper and lower fields at $t_m$, such as $\phi^{\pm}_m$, the two fields would be the same, since $\langle \phi^-_{m},t_{m} |\phi^+_{m},t_{m}\rangle=\delta(\phi_m^+-\phi_m^-)$.
Therefore, we can always integrate out one of them, and this leads us back to the single field insertion at $t_m$.
}

While $\phi^\pm$ prove useful in the practical numerical calculation, another basis of fields more easily reveals the separation of variables, and so we introduce the Keldysh basis 
\begin{align}\label{eq:phicl_phiq}
\phi^{cl}&=\frac{1}{2}\left( \phi^++\phi^- \right),\qquad\phi^q=\phi^+-\phi^-,
\end{align}
where we adopt the coefficients of the transformation from \cite{Aarts:1997kp,Greiner:1996dx}, but the name for the fields from \cite{Kamenev:2009}.
It is important to realize that the notation $(\phi^{cl}, \phi^q)$ does not mean $\phi^{q}$ is ``quantum" and $\phi^{cl}$ is ``classical", as they are both required for the full quantum description. The naming convention of $\phi^{cl}$ follows from the behaviour of this field at the critical point of the action, and its role in the classical-statistical approximation to be discussed later.

Having introduced $(\phi^{cl},\;\phi^q)$, we can now split the partition function into two parts: the initial density matrix and the dynamic part of the path integral.
Since it is natural to pair up $\phi^{q}_{j}$ and $\phi^{cl}_{j+1}$ in the $(\phi^{cl},\; \phi^q)$ basis, we attribute all terms that only consist of $\phi^{cl}_0$, $\phi^{q}_0$ and $\phi^{cl}_1$ to the initial density matrix part.
Nevertheless, the dynamic part still contains $\phi^{cl}_0$ and $\phi^{cl}_1$, but {\it not} $\phi^q_0$.
For the free initial density matrix, we can further integrate out the dummy variable $\phi^q_0$ \cite{Mou:2019tck}, and this integral is in fact a  Wigner-Weyl transform of the initial density matrix \cite{Hertzberg:2019wgx}, which for the quantum field theory in momentum space is of a form,
\begin{align}
W\!\left(\phi^{cl},\pi^{cl}\right) \equiv \int {\mathcal D} \phi^{q}\Big\langle \phi^{cl}+\frac{\phi^{q}}{2} \Big | ~\hat\rho ~ \Big| \phi^{cl}-\frac{\phi^{q}}{2} \Big\rangle \exp\left(-\frac{i}{\hbar}\int \frac{\ud^d p}{(2\pi)^d}\phi^{q} \pi^{cl}\right),
\end{align}
with the notation
$\int \ud^d p ~ fg  \equiv\int \ud^d p ~ \left(f(p)\right)^* g(p)$,
provided that both $f(x)$ and $g(x)$ in the configuration space are real functions.
In practice, the splitting proceeds as 
{\small
\begin{align}
\label{eq:SK_path_integral}
&Z=\int {\mathcal D} \phi  
\langle\phi^+_0;t_0|\hat\rho|\phi^-_0;t_0\rangle e^{ \frac{i}{\hbar}\int_{\cal C}\ud t\;L }
\nonumber \\
=&\!
\!\int\!\! {\mathcal D} \phi_0^{cl} {\mathcal D} \phi_0^{q}  {\mathcal D} \phi_1^{cl}\!\Big\langle \! \phi_0^{cl}\! \! +\!\! \frac{\phi_0^{q}}{2} \Big | \hat\rho  \Big| \phi_0^{cl}\! \! - \! \! \frac{\phi_0^{q}}{2} \!\Big\rangle \exp\!\left(\!-\frac{i}{\hbar}\! \int\!\! \frac{\ud^d p}{(2\pi)^d}\phi_0^{q}\!\left[ \!\frac{\phi_1^{cl}-\phi_0^{cl}\left(1-\omega_p^2\ud t^2/2\right)}{\ud t}\!\right]\! \right)
\!\!\!\int\!\! \prod_{j=1}^{m-1}\!\! \mathcal{D}\phi_j^{q} \mathcal{D}\phi_{j+1}^{cl} e^{-\AI}
\nonumber \\
=&\int{\mathcal D} \phi_0^{cl}    {\mathcal D}  \phi_1^{cl}~ W\!\left(\phi_0^{cl},\pi_0^{cl}\right) \int\! \prod_{j=1}^{m-1}\mathcal{D}\phi_j^{q} \mathcal{D}\phi_{j+1}^{cl} e^{-\AI}
,\end{align}
}%
where from the first line to the second\footnote{We specified the second line by imposing the theory to be free at $t_0$, that is, considering only the quadratic terms in the potential.
Otherwise we should include extra non-linear terms of $\phi^q_0$.
In particular, we have in mind to turn on the interaction gradually or instantly by varying the coupling constants.
}
 we arrange the terms in the exponent into two parts, according to whether or not the terms contain $\phi_0^q$, and $-\AI$ refers to the part without $\phi_0^q$.
Obviously, the conjugate momentum field $\pi^{cl}_0$ is defined via 
\begin{align}
\pi_0^{cl} \equiv  \frac{\phi_1^{cl}-\phi_0^{cl}\left(1-\omega_p^2\ud t^2/2\right)}{\ud t},
\end{align}
which is also the initial $\dot\phi_0^{cl}$, up to the order of ${\mathcal O}\left((\ud t)^2\right)$.

For one thing, we find that the Wigner function of the initial density matrix for the free thermal state reads, up to an overall constant factor, 
\begin{align}
W\!\left(\phi_0^{cl},\pi_0^{cl}\right) =\exp\left(-\frac{1}{\hbar}\int\frac{\ud^dp}{(2\pi)^d}\left[
\frac{\omega_p}{2n_p+1}\left|\phi_0^{cl}\right|^2
+\frac{1}{\omega_p(2n_p+1)}\left|\pi_0^{cl}\right|^2
\right] \right),
\end{align}
where the particle number $n_p$ follows the Bose-Einstein distribution, $n_p=1/(e^{\hbar \omega_p \beta}-1)$, and the frequency $\omega_p$ refers to the frequency of the free theory, namely $\omega_p=\sqrt{p^2+m^2}$.
Concretely, the initialization satisfies,
\begin{align}
\label{eq:initial}
\langle \phi_0^{cl}(p) \Big(\phi_0^{cl}(p')\Big)^\dagger\rangle = \frac{\hbar}{\omega_p}\left(n_p+\frac{1}{2}\right)(2\pi)^d\delta^d(p-p'),
\nonumber \\
\langle \pi_0^{cl}(p) \Big(\pi_0^{cl}(p')\Big)^\dagger\rangle = \omega_p\hbar\left(n_p+\frac{1}{2}\right)(2\pi)^d\delta^d(p-p'),
\end{align}
according to which we can draw random numbers for practical simulations as the initialization.
In particular, we are going to use $\tilde\phi_0^{cl}$ and $\tilde\phi_1^{cl}$ to denote those random numbers generated by this way, and as a convention of the paper, the fields with tildes always refer to the classical trajectories satisfying the classical equation of motion, except of course $\tilde\phi_{0}^{cl}$ and $\tilde\phi_{1}^{cl}$, which serve as the initialization satisfying (\ref{eq:initial}).

For the dynamic part of the path integral, the integrals over all field variables not belonging to the initial condition, there are a number of different ways to express the exponent, depending on which basis we use. In order to better see the classical-statistical approximation, we can use the basis introduced in (\ref{eq:phicl_phiq}), 
the $(\phi^{cl},~\phi^q)$ basis, for which we find 
\begin{align}\label{eq:I_in_cl_q_basis}
 \AI =\left(\frac{-i}{\hbar }\right)\int \ud^d x \sum_{j=1}^{m-1} \left(
\frac{\overline{\overline{\phi^{cl}_{j+1}}}-\phi^{cl}_{j+1}}{\ud t}\phi^q_{j}
+\frac{g_3\ud t}{4}\left(\phi_{j}^q\right)^3 - \frac{\lambda \ud t}{24} {\phi}_{j}^{cl} \left(\phi_{j}^q\right)^3
\right).
\end{align}
Here we have defined 
\begin{align}
\overline{\overline{\phi_{j+1}^{cl}}} \equiv
 2\phi_{j}^{cl}-\phi_{j-1}^{cl}
-\ud t^2
\left(
-\nabla^2 \phi^{cl}_j
+m^2\phi^{cl}_j
-3g_3(\phi^{cl}_j)^2
-\frac{\lambda}{6}(\phi^{cl}_j)^3
\right)
\end{align}
in order to highlight the connection with the equation of motion.
This expression makes it manifest that the exponent $\AI$ is an odd function of $\phi^q$, and we will call those non-linear terms of $\phi^q$ 
the quantum vertices.
The classical-statistical approximation then states that we drop the quantum vertices in (\ref{eq:I_in_cl_q_basis}), in which case we may perform the ${\cal D}\phi^q$ part of the path integral, which leaves us with a delta function $\sim\delta( \overline{\overline{\phi_{j}^{cl}}}(x)-\phi^{cl}_{j}(x) )$ for $j>1$, while we still need to perform the path integral over the initial condition $\phi^{cl}_0$ and $\phi^{cl}_1$. In practise, this means that $\phi^{cl}$ satisfies the classical equations of motion, where the initial data, $\phi^{cl}_0$ and $\phi^{cl}_1$, is drawn from the Gaussian distribution (\ref{eq:initial}). This approximation has been used in many calculations, including early-Universe simulations (for instance \cite{Rajantie:2000nj,GarciaBellido:2002aj,Smit:2002yg,Mou:2015aia}), bubble nucleation \cite{Braden:2018tky,Hertzberg:2019wgx,Blanco-Pillado:2019xny} and thermalization \cite{Aarts:1997kp, Berges:2000ur,Arrizabalaga:2004iw}.

It is clear from these expressions that the exponent $\AI$ is purely imaginary, and so the exponential term in (\ref{eq:SK_path_integral}) is purely a phase, leading to a highly oscillatory integral. Dealing with this oscillatory behaviour is the main role of Picard-Lefschetz thimbles \cite{Witten:2010cx,Cristoforetti:2012su}, and in this paper we use the generalized thimbles of \cite{Alexandru:2017czx,Alexandru:2018fqp,Alexandru:2018ngw} to compute the integral.

\subsection{Lefschetz thimble and Generalized Thimble Method }
\label{sec:thimble}
The central idea of thimble methods is to start with an integration of some function, $e^{-{\cal I}(\underline X)}$, over real variables, $X_j$, and then promote the $X_j$ to complex variables, $Z_j$, whilst also promoting the integrand to a holomorphic function $e^{-{\cal I}(\underline Z)}$\cite{Witten:2010cx,Mou:2019tck,Cristoforetti:2012su,Alexandru:2017czx,Alexandru:2018fqp,Alexandru:2018ngw,Alexandru:2015xva,Alexandru:2015sua,Alexandru:2016gsd,Alexandru:2017lqr,Alexandru:2017oyw,Cristoforetti:2013wha,Mukherjee:2013aga,Aarts:2013fpa,Behtash:2015kna,Cherman:2014sba,Tanizaki:2015rda,Fukuma:2017fjq,Tanizaki:2017yow,Tanizaki:2014xba}. The problem then corresponds to integrating along the surface $\mathbb{R}^n\subset\mathbb{C}^n$ given by Im$(Z_j)=0$, but this surface may be deformed\footnote{The curve must not cross any poles, and the integrand is assumed to vanish asymptotically on the deformed curved.} to some $n-$dimensional surface ${\cal M}\subset\mathbb{C}^n$ without affecting the value of the integral, due to Cauchy's theorem for contour integrals in the complex plane. It is then convenient to perform a coordinate transformation from this curve back to the initial real surface, parametrized by the $X_j$
\begin{align}
\int_{{\mathbb R}^n} \prod_{j=1}^n \ud X_j e^{-{\mathcal I}(\underline X)} =
\int_{{\mathcal M}}\prod_{j=1}^n\ud Z_j e^{-{\mathcal I}( \underline Z)}=
\int_{{\mathbb R}^n}\prod_{j=1}^n\ud X_j {\rm det}\left(\frac{\partial  Z}{\partial  X}\right)e^{-{\mathcal I}( \underline Z( \underline X))}.
\end{align}
The trick to making the integrals converge is to pick an appropriate surface ${\cal M}$, and that is done with the aid of the Picard-Lefschetz flow equations, 
\begin{align}
\frac{\ud  Z_j}{\ud \tau} = \overline{\frac{\partial \AI}{\partial  Z_j}},
\quad\quad
\frac{\ud}{\ud\tau}\left(\frac{\partial  Z_j}{\partial  X_k}\right) = \overline{\frac{\partial^2 {\mathcal I}}{\partial  Z_j\partial  Z_l}\frac{\partial  Z_l}{\partial  X_k}},
\end{align}
where the second equation can be derived from the first.
As initial conditions for the flow, one takes $Z_j=X_j$,  and sets $J_{jk}=\partial  Z_j/\partial  X_k$ to be the $n\times n$ identity matrix. In this way, each point $\underline X$ on the real surface is flowed into the complex plane. The surface $\mathcal{M}$ is called a generalized thimble \cite{Alexandru:2017czx,Alexandru:2018fqp,Alexandru:2018ngw}, and its shape depends on how long one chooses to flow the equations. In the limit of infinite flow time one recovers the Lefschetz thimbles.

In the case of field theory, our $\phi_i$ in (\ref{eq:SK_path_integral}) are promoted to complex variables, in analogy to the $X_j\to Z_j$ above.\footnote{To avoid confusion: The time variable indexing the fields $\phi$ lives in a complex plane on the Keldysh contour. The time index is part of the $j$ label on $X_j$. But now the domain of integration, the values taken by the field variables $\phi_j$, is also deformed into the complex plane.} Along the flow, ${\rm Re}[\AI]$ always increases, so the weight $e^{-{\rm Re}[\AI]}$ decreases exponentially, making the integrand of (\ref{eq:SK_path_integral}) take the form of damped oscillations, leading to a well-behaved integral.
In practice, it is convenient to combine this with the determinant of the Jacobian and construct a probability distribution function $P=e^{-{\rm Re}[{\mathcal I}]+\ln |{\rm det}(J)|}$ for use in the Monte Carlo evaluation of the integral \cite{Alexandru:2017lqr,Mou:2019tck}.
Now, by generating samples according to this probability distribution function, we are able to compute the expectation value of any observable, which is done by reweighting the observable by $e^{-i{\rm Im}[{\mathcal I}]+i{\rm arg}\left({\rm det}(J)\right)}$, as this is what remains of the $e^{-\AI}$ after we have accounted for $P$.

For a single initialization, i.e. one choice of $\tilde\phi^{cl}_0$ and $\tilde\phi^{cl}_1$, we can measure the expectation value due to its single generalized thimble as
\begin{align*}
\langle {\mathcal O}\rangle_{\rm GT} &=
\frac{\Big\langle e^{-i{\rm Im}[{\mathcal I}]+i{\rm arg}\left({\rm det}(J)\right)} {\mathcal O}\Big\rangle_P }{\Big\langle e^{-i{\rm Im}[{\mathcal I}]+i{\rm arg}\left({\rm det}(J)\right)}\Big\rangle_P }
,
\end{align*}
where the brackets $\langle...\rangle_P$ correspond to an average using $P$ as the probability distribution function. To compute the full quantum average, denoted $\langle...\rangle$, we need to average over an ensemble of initializations  $\tilde\phi^{cl}_0$, $\tilde\phi^{cl}_1$ drawn from a Gaussian as determined by the initial distribution (\ref{eq:initial}) \cite{Mou:2019tck}.

\subsection{Critical points}
\label{sec:criiticalpoints}

Of central importance to our discussion are the critical points (or critical configurations), the fixed points of the flow, $\partial \AI/\partial Z=0$. These are the classical trajectories, and so as the manifold $\mathcal{M}$ is deformed by the flow, it does so while the classical trajectories stay pinned at their real values. All configurations away from the critical points give an exponentially suppressed contribution (depending on how much we flow), while all the critical points contribute with equal weight. Hence it is essential to identify and include all critical points, and their neighbourhoods in configuration space.

By definition, we only need to solve 
\begin{align}
\frac{\partial \AI}{\partial \phi_{j}^{q}(x)}=0 
,\quad {\rm and}\quad
\frac{\partial \AI}{\partial \phi_{j+1}^{cl}(x)}=0
,\quad {\rm for~all~}x{\rm~and~}  j\geq 1,
\end{align}
with the explicit expression of $\AI$ in (\ref{eq:I_in_cl_q_basis}).
The linear dependence on $\phi_m$, which is treated as a $\phi^{cl}$ field, together with the odd dependence on $\phi^q$,  allows us to solve the equations exactly.
Firstly, since $\phi_m(x)$ only couples to $\phi_{m-1}^{q}(x)$, we obtain $\phi_{m-1}^q(x)=0$ as the solution of $\partial \AI/\partial \phi_{m}(x)=0$.
Then, given $\phi_{m-1}^q(x)=0$ for all $x$, we can derive $\phi_{m-2}^q(x)=0$ from $\partial \AI/\partial \phi_{m-1}^{cl}(x)=0$.
In fact, we can continue this process to discover that all $\phi^q(x)=0$ at the critical points.
With this in mind, $\partial \AI/\partial \phi_{j}^q(x)=0$ leads to the classical equation of motion for $j\geq 1$,
\begin{align}
\tilde\phi_{j+1}^{cl} 
- 2\tilde \phi_{j}^{cl}+\tilde \phi_{j-1}^{cl}
+\ud t^2
\left(
-\nabla^2 \tilde \phi^{cl}_j
+m^2\tilde \phi^{cl}_j
-3g_3(\tilde \phi^{cl}_j)^2
-\frac{\lambda}{6}(\tilde \phi^{cl}_j)^3
\right)
=0,
\end{align}
where we use tilde to refer to the solution of $\phi^{cl}$, given the initialization $\tilde\phi_0^{cl}(x)$ and $\tilde\phi_1^{cl}(x)$. Unsurprisingly, we have shown that the stationary points are the classical trajectories $\tilde{\phi}^{cl}$. And by sampling the entire ensemble of initial conditions $\tilde\phi_0^{cl}(x)$, $\tilde\phi_1^{cl}(x)$ we can ensure that we include all of them. 

For a given initial condition $\tilde\phi_0^{cl}(x)$, $\tilde\phi_1^{cl}(x)$, we then have a unique classical trajectory, and we proceed to generate a Markov chain of random trajectories starting from there. The resulting trajectories are then, in general, no longer fixed points of the flow, and are therefore evolved into some complex-valued trajectory through applying the flow equation. The thimble corresponding to the original classical trajectory is the collection of all such nearby trajectories flowed into the complex space. 

Still, had all the field variables $\phi_j$ been part of this Monte-Carlo sampling, one could imagine at some stage randomly generating another classical trajectory. This is then the fixed point corresponding to another neighbourhood of trajectories, belonging to another thimble than the original one. This means that when sampling all the field variables simultaneously, one should take care to properly include all the thimbles, in effect finding all the neighbourhoods of classical trajectories. 

In our case, we sample the initial conditions separately, and keep them fixed, while the Monte-Carlo sampling of the other variables takes place. 
Thus as an initial value problem, the existence and uniqueness of the critical point in the complex space are guaranteed.
This means that we will never at random end up on a different classical trajectory, because there is only one classical trajectory corresponding to given values of $\tilde\phi_0^{cl}(x)$, $\tilde\phi_1^{cl}(x)$ (or indeed the field variables at any two times, we choose to keep fixed). And so in our two-step sampling, we are sure to find all thimbles and that in any Markov chain, only one thimble is present.

We stress that the conclusion here applies to the quantum field theory with {\em any} number of spatial dimensions.

In summary, our approach has a number of advantages for real-time simulations using the Keldysh contour:
It admits the two-part splitting into an ``external" average over realisations of the initial density matrix and the ``internal" Monte-Carlo sampling of the rest of the field variables in the path integral.
We can generate the initialization according to the initial density matrix, and with each external initialization there exists one and only one critical point/thimble for the internal part.
Thirdly, the critical point $(\phi^q=0,~\phi^{cl}=\tilde\phi^{cl})$ lies in the real plane, as long as $\tilde\phi_0^{cl}$ and $\tilde\phi_1^{cl}$ are real, which is the case in our approach. Thus the Lefschetz thimble and the generalized thimble coincide at the unique critical point.
At the moment, it is easier to implement an algorithm to generate samples on the generalized thimble than on the Lefschetz thimble, so we are going to adopt the Generalized Thimble Method in the following calculation.
For more details and numerical applications, see \cite{Mou:2019tck}.

\section{Tunnel splitting of the ground state}
\label{sec:tunnel}

Having set up our formalism and algorithm, we wish to apply it to the inherently quantum mechanical process of tunneling. We therefore introduce a system with a double well potential,
\begin{align}\label{eq:potential:}
{\mathcal V}=\frac{1}{2}\phi^2\left(1-g\phi\right)^2
,
\end{align}
which corresponds to 
$m=1
,~
g_3=g
,~
\lambda=12g^2$ in (\ref{eq:V}), and we shall take $\hbar=1$ in the simulations. In the following, we will consider field theory in zero spatial dimensions, i.e. quantum mechanics.

The two minima of the potential are situated at $\phi=0$ and $\phi=1/g$ respectively, and the height of the barrier between the two minima is $1/(32g^2)$.
We set the system up with an initial Gaussian vacuum state of the $\phi=0$ minimum.
When $g$ is small, the separation between the two minima is large and the barrier is high, but the wave function will nevertheless oscillate between the two minima.
This is a textbook problem \cite{Landau,Zinn} known as tunnel splitting of the ground state, which is why we use it as a test of the thimble approach.

To get an idea of how the system behaves we may start off by making the approximation of focussing only on the first two energy eigenstates.
These are approximately given by the superposition of the two Gaussian vacuum states located around each minimum,
\begin{align}
\Psi_+\simeq\frac{\Psi_L+\Psi_R}{\sqrt{2}}
,\quad
\Psi_-\simeq\frac{\Psi_L-\Psi_R}{\sqrt{2}},
\end{align}
where $\Psi_L$ is the Gaussian wavefunction associated to the $\phi=0$ minimum, and $\Psi_R$ is the Gaussian wavefunction associated to the $\phi=1/g$ minimum.

Thus if we take the initial state to be $\Psi_L$, we expect the wave function to evolve as,
\begin{align}\label{eq:two_state_approx}
\Psi(t)\simeq\frac{e^{-iE_+t}\Psi_++e^{-iE_-t}\Psi_-}{\sqrt{2}}.
\end{align}
In particular, if we measure the average $\phi$, we would obtain
\begin{align}\label{eq:approx_evolution}
\int \ud \phi\; \Psi^*(t)\phi\Psi(t)
&\simeq\int \ud \phi \;\frac{1}{2}
\left(\Psi_+\phi\Psi_++\Psi_-\phi\Psi_-+e^{-i\Delta Et}\Psi_+\phi\Psi_-+e^{i\Delta Et}\Psi_-\phi\Psi_+\right)
\nonumber \\
&\simeq\int \ud \phi\; \frac{1}{2}
\left(\Psi_L\phi\Psi_L+\Psi_R\phi\Psi_R+\cos(\Delta Et)\left(\Psi_L\phi\Psi_L-\Psi_R\phi\Psi_R\right)\right)
\nonumber \\
&\simeq\frac{1}{2g}
\left[1-\cos(\Delta Et)\right]
,
\end{align}
where, for last relation, we utilize the fact that $\int \ud \phi \Psi_L\phi\Psi_L=0$ and $\int \ud \phi \Psi_R\phi\Psi_R=1/g$.

This shows that the oscillation is described by a single parameter $\Delta E$, which is the energy difference between the first two energy eigenstates.
In the literature, there exist many ways to compute the energy difference, here we mention three of them:

\vspace{0.2cm}
\noindent\textbf{WKB estimate of $\Delta E$:}\\
The textbook of Landau and Lifshitz \cite{Landau} provides a WKB analysis of the problem, leading to the formula,
\begin{align}\label{eq:WKB_example}
\Delta E = \frac{1}{\pi} \exp\left[-\int_{a}^{b} \sqrt{2\left(V-E_0\right)} \ud \phi \right]
.
\end{align}
where $a$ and $b$ are turning points, determined by $V-E_0=0$. In theory, $E_0$ should be the average energy between the first two states, but this may be no easier to calculate than $\Delta E$. However, in practice we may take the free vacuum energy, $0.5$, to be an estimate of $E_0$. As $g$ becomes larger, $g>0.25$ in our example, then the height of the barrier is lower than $0.5$, meaning that the WKB formula (\ref{eq:WKB_example}) is no longer valid, so we can only trust this approximation for $g\ll 0.25$.

\vspace{0.2cm}
\noindent\textbf{Instanton estimate of $\Delta E$:}\\
The tunnelling property can also be captured by an instanton calculation, and the energy difference is estimated via a simple formula \cite{Zinn,Garg},
\begin{align}\label{eq:instanton}
\Delta E = \frac{2}{g\sqrt{\pi}} \exp\left(-\frac{1}{6g^2}\right)
.
\end{align}
The formula holds when $g$ is small and we note that one does not need to know the average energy $E_0$ {\it a priori}.

\vspace{0.2cm}
\noindent\textbf{Finite Fock space estimate of $\Delta E$:}\\
Finally, for this simple problem, we can find $\Delta E$ to high precision by approximating the Fock space with a finite set of vectors \cite{Matrix}
\begin{align}
|n\rangle = \frac{1}{\sqrt{n!}}(a^\dagger)^n |0\rangle
,\quad
a^\dagger | n \rangle = \sqrt{n+1} | n+1 \rangle
,\quad
a | n \rangle = \sqrt{n} | n-1 \rangle
,
\end{align}
and write $\phi=(a^\dagger +a)/\sqrt{2}$ into a matrix,
\begin{align}
\phi=\left(\begin{array}{cccc}
0 & 1/\sqrt{2} & 0  & \cdots \\
1/\sqrt{2} & 0  & 1 & \cdots \\
0 & 1 & 0 & \cdots \\
\vdots &\vdots & \vdots& \ddots 
\end{array}\right)
.
\end{align}
The lowest two eigenvalues of $H=H_0-g\phi^3+g^2\phi^4/2$ are then the energy of the ground state and the first excited state, respectively.
With a $100\times 100$ matrix, we can already achieve a high precision, as will be seen in the next section.

\begin{table}[H]
\centering
\begin{tabular}{ |c|c|c|c|  }
 \hline
    & $g=0.2$    & g=0.3 &   g=0.5\\
 \hline
WKB & \begin{tabular}[x]{@{}c@{}}$\Delta E = 0.087$ with $E_0=0.5$~\end{tabular} & - & -\\
 \hline
Instanton & $\Delta E = 0.087$  & $\Delta E = 0.590$  & $\Delta E = 1.159$ \\
 \hline
Matrix & $\Delta E = 0.061 $ & $\Delta E =0.286 $ & $\Delta E =0.637 $ \\
 \hline
\end{tabular}
\caption{
Both WKB and Instanton approximations to $\Delta E$ work best when $g$ is small.
In the cases of $g=0.3$ and $0.5$, $E_0$ is already larger than the height of potential barrier, so the WKB approach is not valid.
}
\label{tab:deltaE}
\end{table}
In \cref{tab:deltaE}, we present $\Delta E$ for different $g$ using the three different methods. 

Having gained an analytic understanding for the behaviour of the wavefunction, specifically (\ref{eq:approx_evolution}), we may now solve the full Schr\"odinger equation numerically, and use our analytic approach to interpret the results.

\subsection{Solving Schr\"odinger equation}
\label{sec:Schrodinger}

\begin{figure}[t]
\hspace{1cm}
\includegraphics[width=0.8\textwidth]{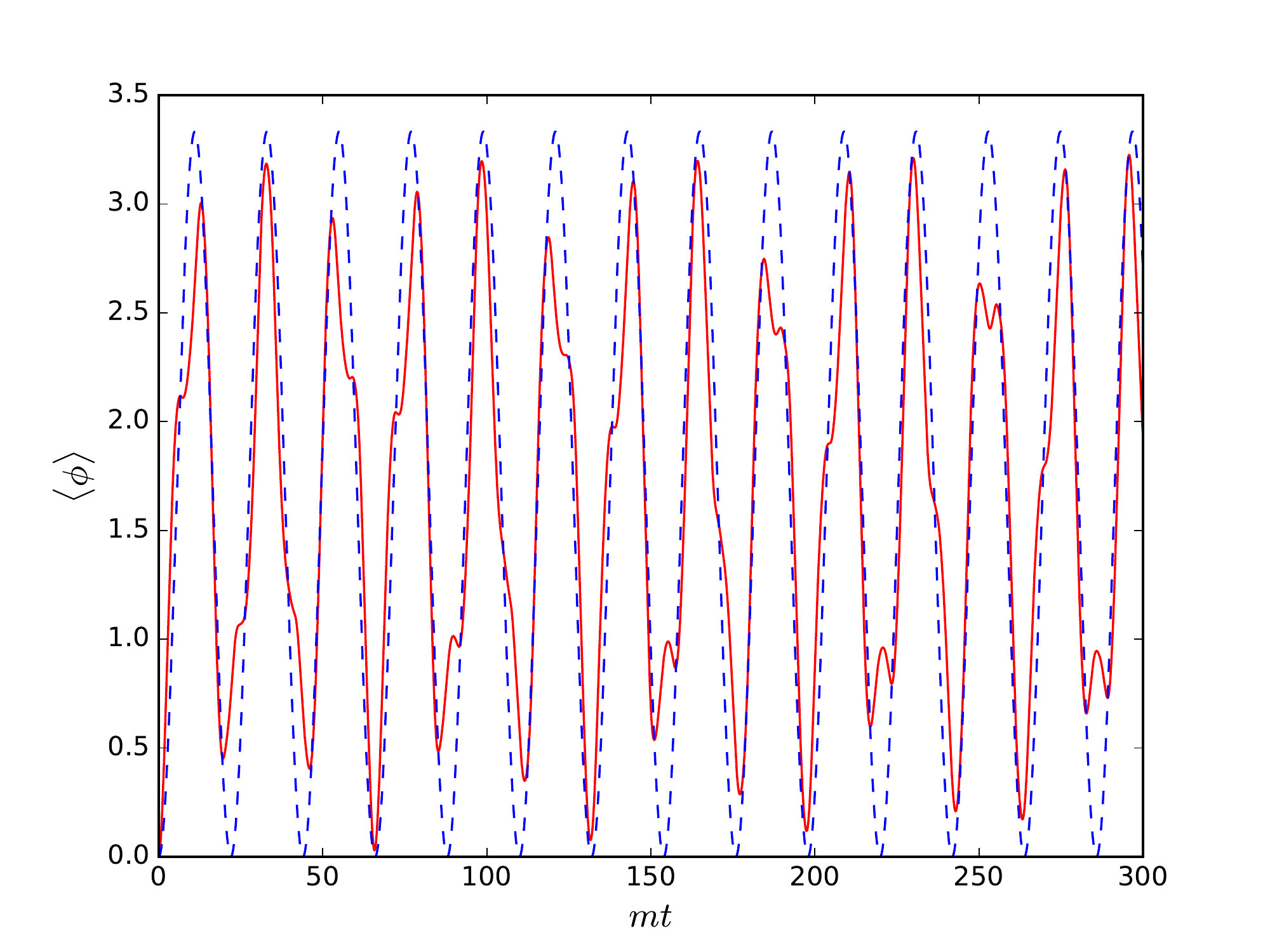}
\caption{
Expectation value $\langle \phi \rangle$ for the potential (\ref{eq:potential:}).
The dashed blue line refers to the fit with function $\frac{1}{2g}(1-\cos(\Delta E t))$, with $g=0.3$ and $\Delta E = 0.285763$, matching the matrix approximation found in Tab. \ref{tab:deltaE}.
Also notice there exist oscillations with higher frequencies, which indicate that other eigenstates contribute too, but they are not as dominant as the first two eigenstates.
}
\label{fig:fit}
\end{figure}

\begin{figure}[H]
\begin{tabular}{lr}
\hspace{-3cm}
\includegraphics[width=0.65\textwidth]{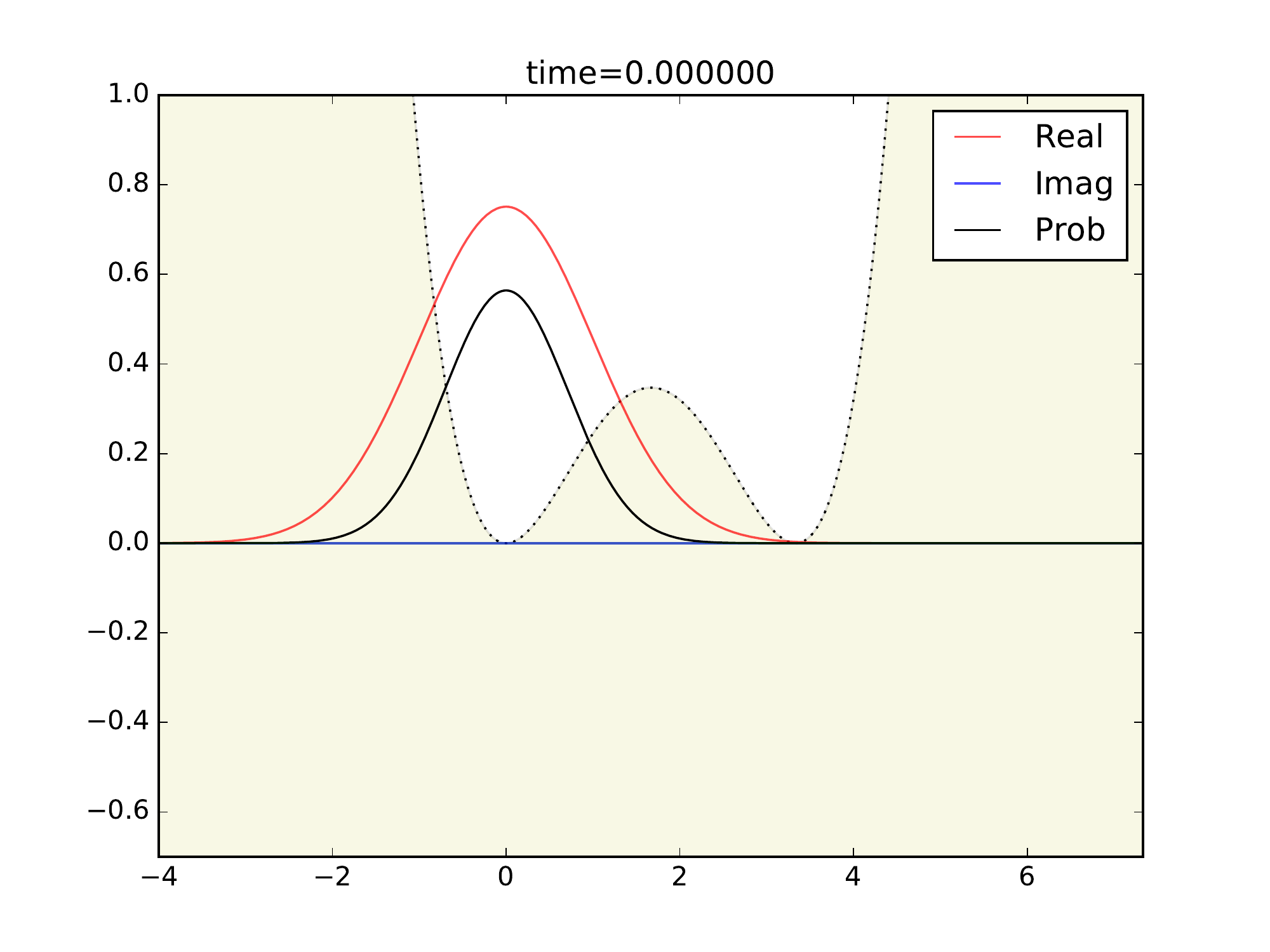} & \hspace{-1cm}
\includegraphics[width=0.65\textwidth]{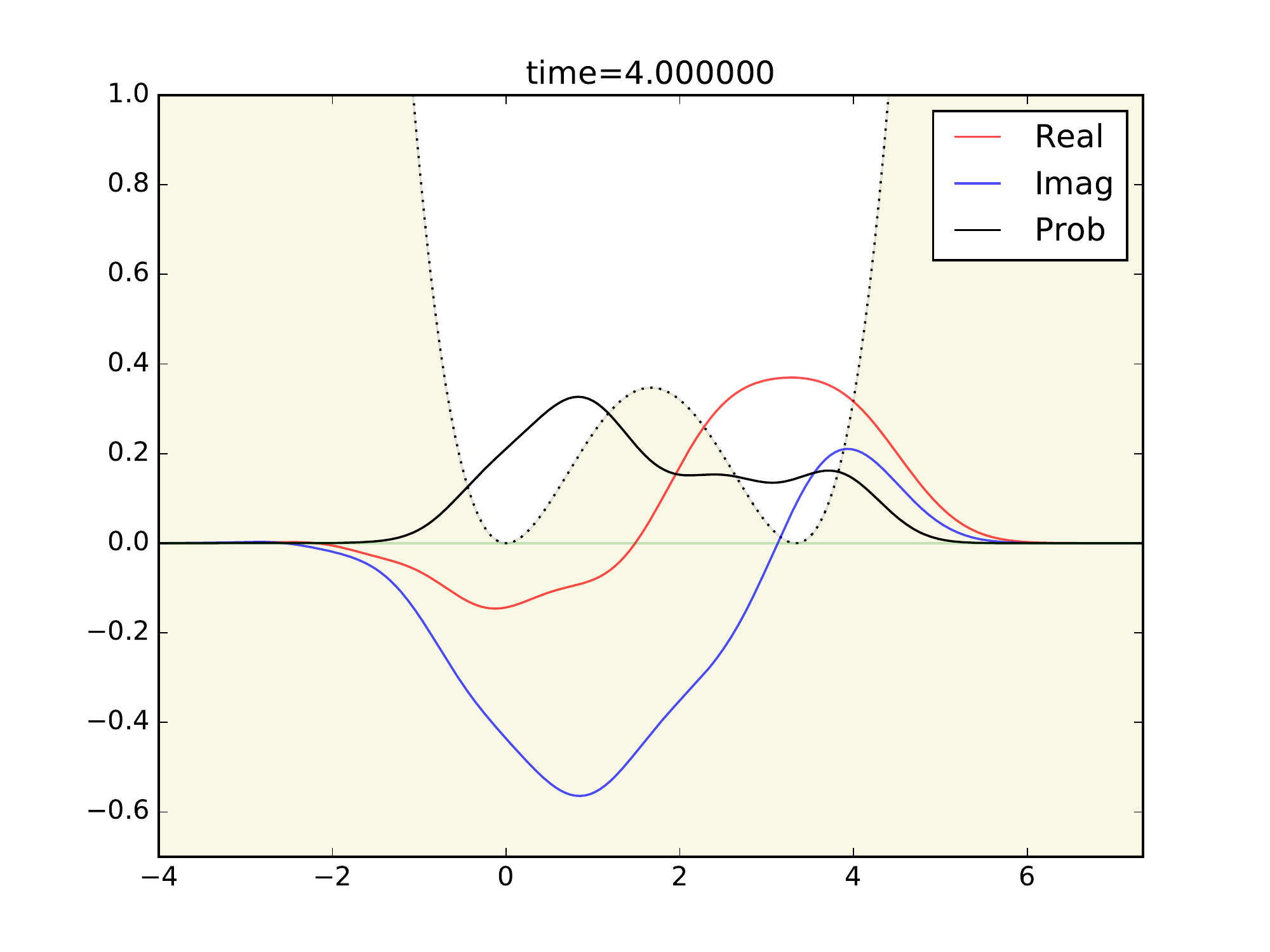} 
\end{tabular}
\caption{
Plots of the real and imaginary parts of $\Psi$ and the probability $\Psi\Psi^*$ at times $mt=0,\;4$ for $g=0.3$.
}
\label{fig:snapshot}
\end{figure}

The time dependent Schr\"odinger equation reads,
\begin{align}
i\hbar \frac{\partial  }{\partial t} \Psi = -\frac{\hbar^2}{2} \frac{\partial^2}{\partial \phi^2 } \Psi + V\Psi
,
\end{align}
and we give the wave function an initial condition associated to the Gaussian ground state of the $\phi=0$ minimum,
\begin{align}
\Psi(\phi, 0)=\left(\frac{m}{\hbar \pi}\right)^{1/4}\exp\left(-\frac{m\phi^2}{2\hbar}\right)
.
\end{align}
It is fairly straightforward to simulate this system with such an initial condition, and for an explicit realization with Python, see \cite{python}. In \cref{fig:fit}, we show the expectation value of $\phi$ in the time range $0<mt\leq300$, with $g=0.3$, as calculated with the full Schr\"odinger equation (solid red curve), and the analytic approximation (\ref{eq:approx_evolution}) based on the ansatz of (\ref{eq:two_state_approx}). The approximate solution shows good agreement to the full solution, with the oscillations clearly visible at the expected frequency, along with further structure that is not captured by the two-state approximation. The presence of this extra structure is seen by looking at a snapshot of the wavefunction at some later time, as  in \cref{fig:snapshot}

Given that the two-state approximation yields good agreement with the full solution we may introduce the idea of the tunnelling time, given by $t_{tun}=\frac{\pi}{\Delta E}$, and we see from Tab. \ref{tab:deltaE} that small $g$ leads to longer tunnelling times, as is to be expected given that in this regime the barrier height is increased, and the vacua are further separated. We should note here that one can in fact rescale $\phi$ to remove $g$, at the cost of redefining $\hbar\to g^2\hbar$. Here we would interpret the small $g$ limit as a small $\hbar$ limit, and again we expect the tunnelling times to be increased in this limit.

\subsection{Results with the thimble approach}
\label{sec:result}

We adopt the algorithm used in \cite{Mou:2019tck} to implement the Markov Chain Monte Carlo evaluation of the integral.
In practice, we first generate initial $\tilde\phi^{cl}_0$ and $\tilde\phi^{cl}_1$ according to the Gaussian initial distribution (\ref{eq:initial}).
Then, for each initialization, we utilize a Metropolis–--Hastings algorithm to generate samples on the generalized thimble.
Specifically, according to the proposal distribution, the algorithm is classified as a Random Walk Metropolis–--Hastings. 
A robust choice of the proposal function is $\exp\left(-(\ud x)^T J_n^\dagger J_n \ud x/\delta^2\right)$, given in \cite{Alexandru:2017lqr}.
Notice the proposal function above is expressed in the real space, and it corresponds to $\exp\left(-(\ud z)^\dagger\ud z/\delta^2\right)$ in the complex space, given that $\ud z=J\ud x$.
Therefore the covariance matrix is proportional to the identity matrix in the complex manifold.
Random Walk Metropolis-Hastings algorithm becomes hard to explore the target space, when the number of integration variables is large, and in practice we find that when the total number of variables is larger than $\sim$20, each Markov chain will require ${\mathcal O}(100)$ CPU-hours or more.
Thus, in order to guarantee we achieve ergodicity for each chain, we restrict to 18 field variables, and spend more computational time in increasing the number of updates. 

The numerical challenge we encounter here also reflects the fact that the pure quantum tunnelling effect is hard to simulate, as it is accompanied with exponential suppression of the evolution across the barrier, for instance as in \cite{Mou:2017gnm}.
To have a sensible signal of tunnelling, a reasonably-sized coupling constant is required.
But for the double-well potential, when the quantum tunnelling effect is big, so is the classical-statistical effect, whereby the field has enough energy to classically roll over the barrier for a significant proportion of the random initial conditions $\tilde\phi^{cl}_0$ and $\tilde\phi^{cl}_1$ . 
Therefore, we want the simulations to be able to distinguish the two effects with a reasonable number of field variables.
We shall present results for $g=0.3$ and $0.5$, both of which correspond to $g>0.25$, and so the height of potential barrier is close to, but just below, $0.5$, i.e. the Gaussian vacuum energy. As we start the system in the Gaussian ground state, this means that the average energy of the particle is close to the barrier height, meaning that the classical-statistical approximation is just about able to probe the second minimum. 

\begin{figure}[t]
\begin{tabular}{lr}
\hspace{-2.3cm}
\includegraphics[width=0.65\textwidth]{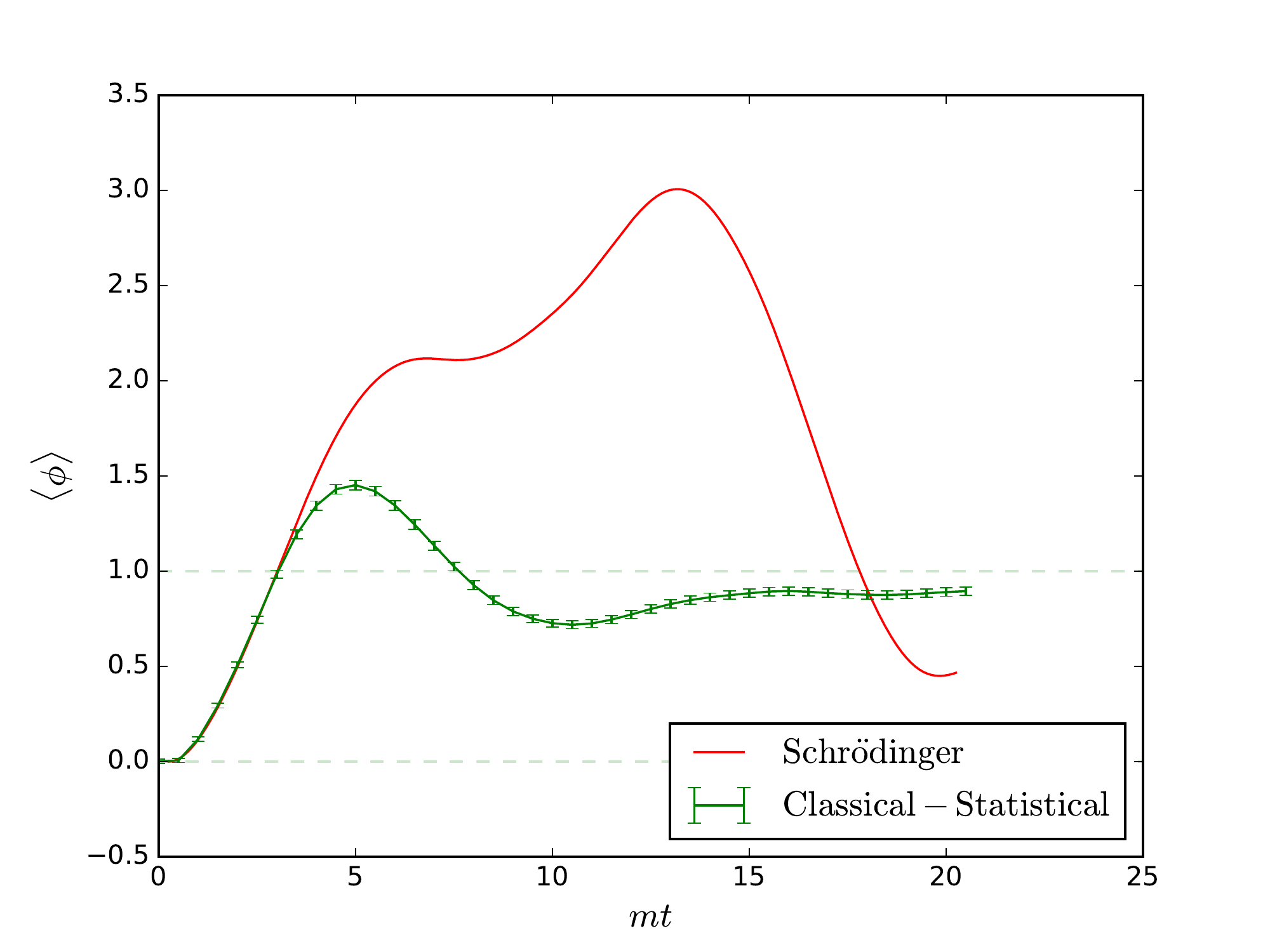} & \hspace{-1.5cm}
\includegraphics[width=0.65\linewidth]{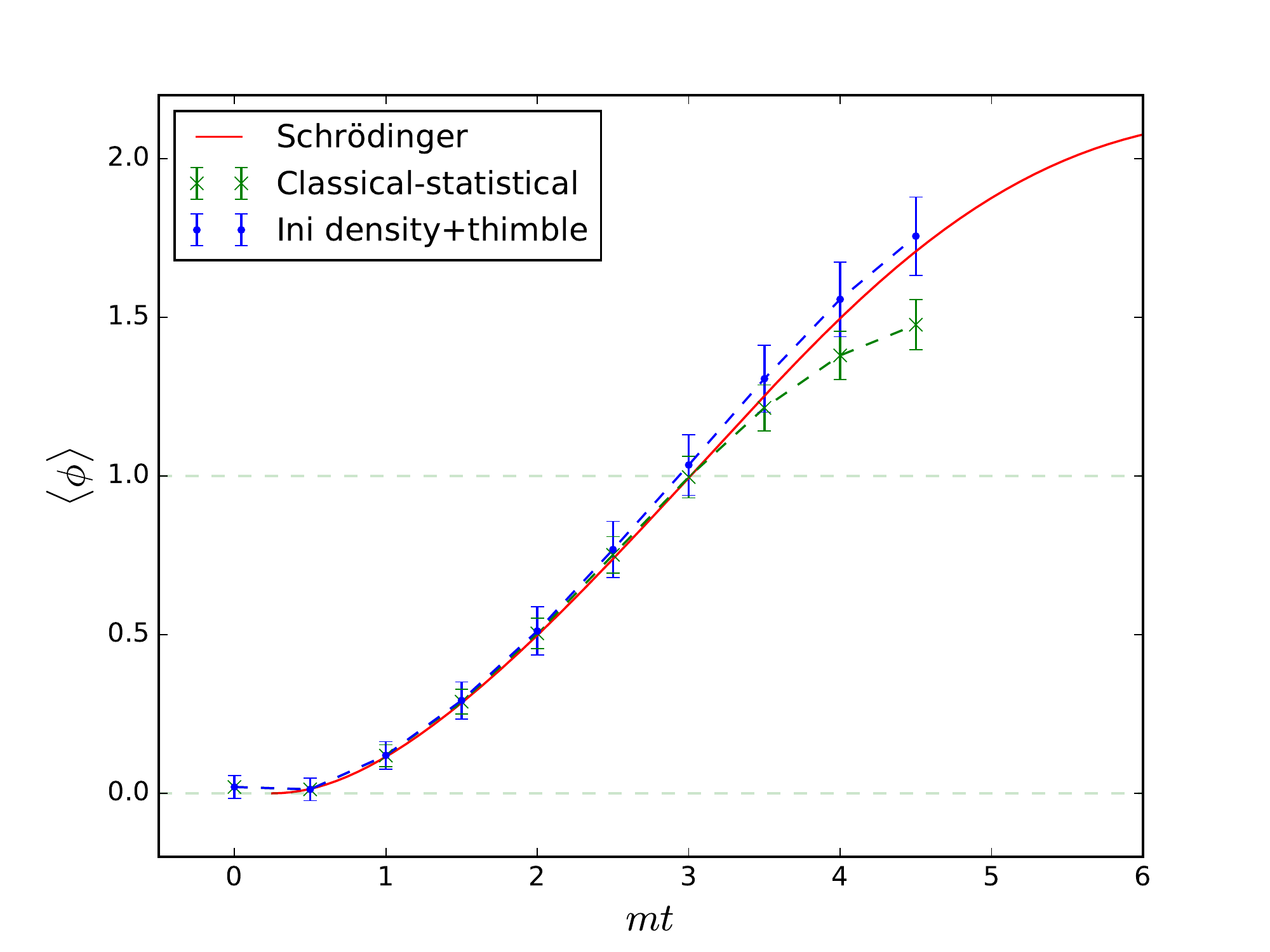}
\end{tabular}
\caption{
Results of the simulations with $g=0.3$ for: full Schr\"odinger evolution (left, solid curve); classical-statistical approximation (left, curve with 1$\sigma$ statistical error bars); a comparison of all three methods, Schr\"odinger, classical-statistical and initial density plus thimble, over a reduced evolution time.
}
\label{fig:g03}
\end{figure}

In \cref{fig:g03} we show the results for $g=0.3$. The plot on the left gives the solution for the full Schr\"odinger equation (solid curve), and the results for the classical-statistical approximation are shown as the curve with 1$\sigma$ statistical error bars. For this value of the coupling the barrier is located at $\phi=5/3$, and it is clear that the classical-statistical approximation starts to break down as the expectation value moves into the second minimum. 
The plot confirms our expectation that the classical-statistical approximation can not capture what really happens during the quantum tunnelling.
The right hand plot of \cref{fig:g03} shows a comparison of the three methods, the path integral and the Schr\"odinger methods are seen to agree within error bars, as they should, while the classical-statistical approximation is starting to deviate at later times.

\begin{figure}[H]
\centering
\includegraphics[width=0.65\textwidth]{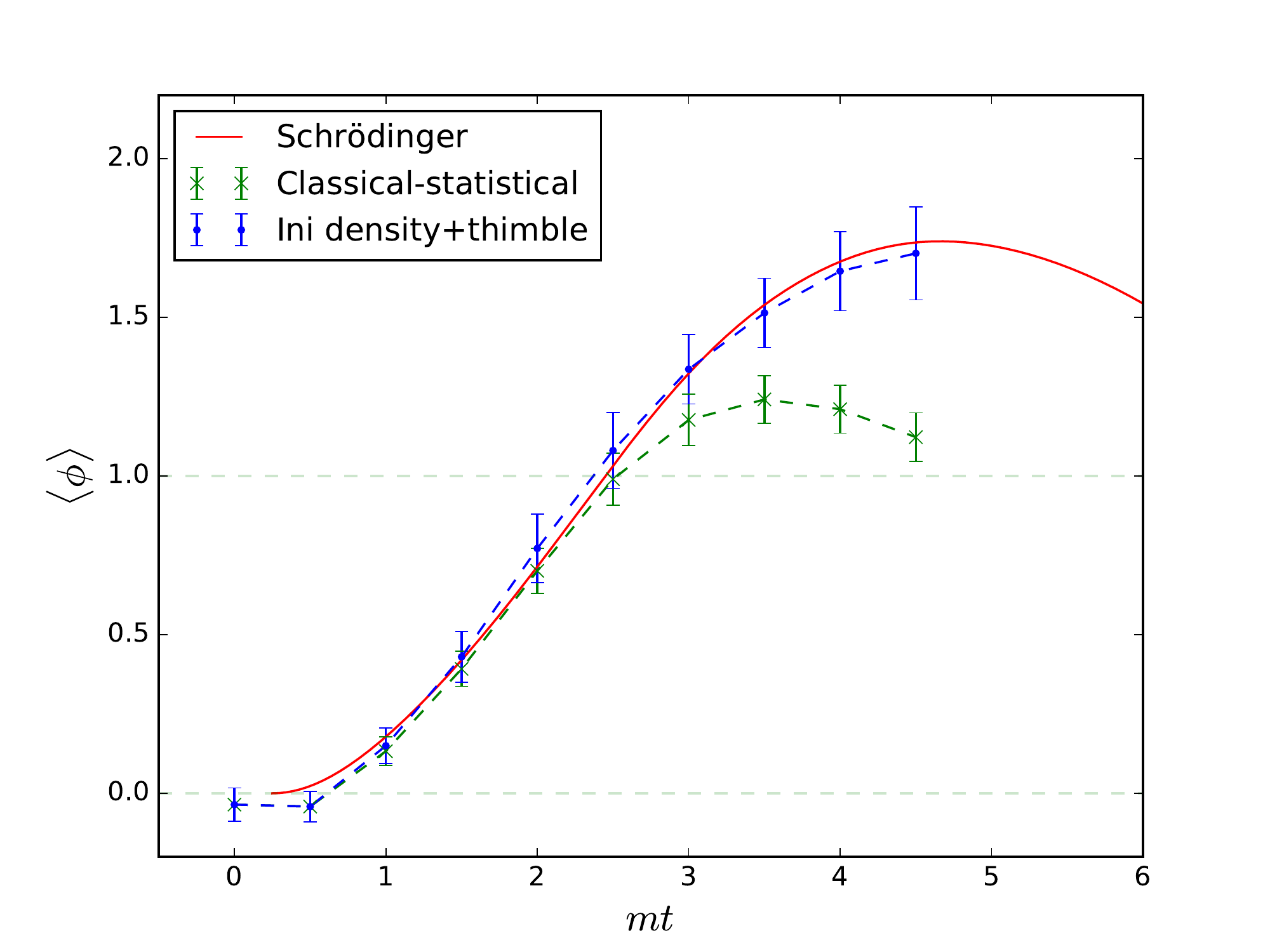}
\caption{
Results of the simulations with $g=0.5$:  a comparison of all three methods, Schr\"odinger, classical-statistical and initial density plus thimble.
}
\label{fig:g05}
\end{figure}
The simulations with a slightly higher coupling, $g=0.5$, correspond to a smaller tunnelling time, and so behave more quantum mechanically. The results of this case are shown in \cref{fig:g05}, where we see that the Schr\"odinger and the thimble approaches again agree within error bars, but now the classical-statistical approximation is showing a more pronounced deviation than in the $g=0.3$ case.
For $g=0.3$, we have implemented 400 initializations, each with 20 millions updates, for $g=0.5$ we used 200 initializations, each with 50 millions updates.

The key difference between the classical-statistical approximation and the full thimble approach is that the approximation only uses the critical point for each set of initial conditions, i.e. the solution to the classical equations of motion for that initial data. The full thimble, however, is able to explore more of the field space in order to get a better estimate for the path integral. 

\begin{figure}[H]
\centering
\centering
\includegraphics[width=0.65\textwidth]{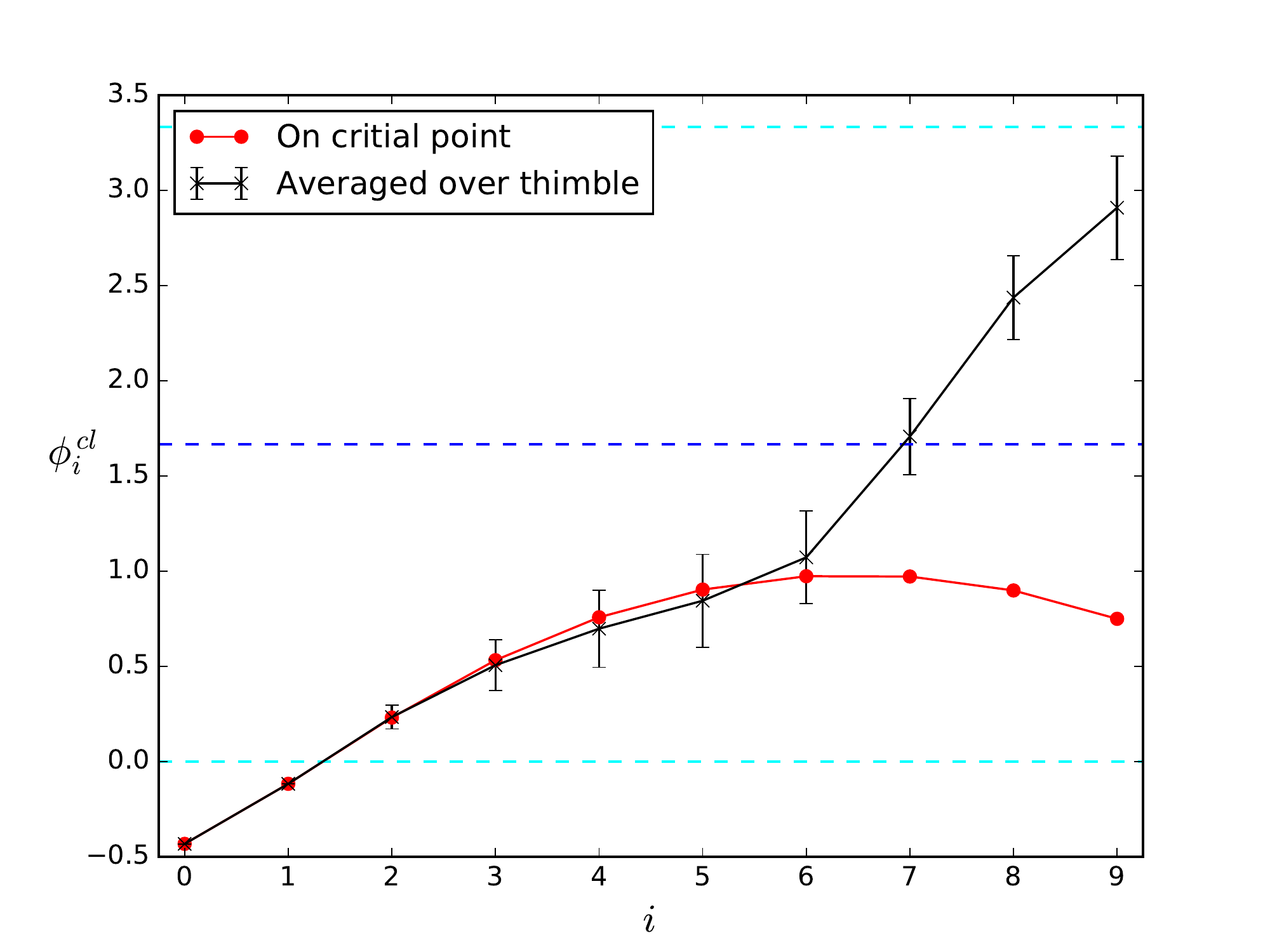}
\caption{
$\phi_i^{cl}$ restricted to a single thimble (single initialization) with $g=0.3$.
The red dots refer to the values of $\phi_i^{cl}$ on the critical point.
So the red line represents the classical trajectory, which is in this case oscillating only around one of the two minima, whose values are denoted in the plot by the dashed cyan lines, while the maximum of the potential by the dashed blue line.
The black error bar indicates the mean and one standard error of ${\rm Re}[\langle \phi_i^{cl}\rangle ]$ evaluated over the whole thimble.
}
\label{fig:hist}
\end{figure}

In our simulations, the thimble is a 16 (real-)dimensional subspace of $\mathbb{C}^{16}$, and so it is difficult to visualize. But we can show a comparison the field values at the critical point versus the average over the whole generalized thimble.
In \cref{fig:hist} we present the data for a single Monte-Carlo chain, for one choice of initial data. This shows the statistics of $\phi_i^{cl}$ for each time slice of the path integral in the $g=0.3$ case, where the tunnelling rate is suppressed by the larger barrier height. The solid red lines show the solution to the classical equations of motion for the same initial data, which will form one member of the ensemble of the classical-statistical approximation. 
We see that at early times the generalized thimble values match the red lines, but deviate at later time slices, leading to the discrepancy between the full quantum behaviour and the classical-statistical approximation.

\section{Conclusion}
\label{sec:conclu}
We have provided a calculation of quantum mechanical tunnelling using the generalized thimble approach \cite{Alexandru:2017czx,Alexandru:2018fqp,Alexandru:2018ngw}, and compared it to the full Schr\"odinger equation computation, as well as the classical-statistical approximation \cite{Aarts:1997kp}. The thimble method presented by the authors in \cite{Mou:2019tck} explicitly breaks up the path integral into initial conditions (external) and a dynamic (internal) part, and so is ideally suited to understanding the classical-statistical approximation, where the path integral is approximated by summing over solutions to the classical equations of motion. We have demonstrated that the thimble calculation of the real-time path integral does indeed reproduce the expected results of the Schr\"odinger equation, and we have seen how it improves on the classical-statistical approximation by exploring more of the field trajectory space required in calculating the path integral.

While the Schr\"odinger equation is by far the superior method for calculating these tunnelling processes in quantum mechanics, the Schr\"odinger functional method is not well-suited to numerical studies of quantum field theory, which is where we expect the thimble approach to be superior in understanding real-time, non-perturbative quantum dynamics of (at least) scalar field theories.

In the present work, we adopt the Generalized Thimble Method, but because of the uniqueness of the thimble for each initialization, it is also possible to sample directly on the Lefschetz thimble.
The latter approach might be optimal when there are more integration variables.
In either case, it becomes more and more difficult to explore the manifold of the thimble through a Markov chain, as the dimension of the manifold increases. Also, it seems that the tunnelling process in itself also amplifies the difficulty.

The numerical challenge is substantial. Given $n$ integration variables, the most time consuming part of solving the flow equations is to calculate ${\rm det}(J)$ which alone requires ${\mathcal O}(n^3)$ computational time (or less, see \cite{Alexandru:2017lqr}). Another large factor is the total number of updates of the Markov chain, whose exact form depends on the algorithm used. To explore the manifold of each thimble, we are now using the Random Walk Metropolis-Hastings algorithm, which on the tunnelling problem behaves less efficiently in higher dimensions.
The hope is that we can find some efficient Monte Carlo algorithm, which can maintain  ${\mathcal O}(n)$ or some other polynomial power. This would mirror the efficiency gained for imaginary time lattice simulations through Hybrid Monte Carlo or Metropolis with local update. An interesting option is to implement a machine learning algorithm to better estimate the thimble \cite{Moss:2019fyi,Albergo:2019eim}. Another is to perhaps adapt a complex Langevin algorithm \cite{Berges:2006xc,Aarts:2013fpa,Aarts:2014nxa} to explore along a thimble rather than in the entire complexified manifold.

\vspace{0.4cm}

{\noindent \bf Acknowledgements:} ~~
PMS is supported by STFC Grant No. ST/L000393/1 and ST/P000703/1, AST and ZGM are supported by a UiS-ToppForsk grant. The authors were also supported by a ECIU travel grant. The numerical work was performed on the Abel supercomputing cluster of the Norwegian computing network Notur.

\end{document}